\documentclass[thmsa,11pt]{article}
\usepackage{amsmath}
\usepackage{amssymb}
\usepackage{amsfonts}
\usepackage{graphicx}
\usepackage{color}
\usepackage[stable]{footmisc}

\numberwithin{equation}{section}

\begin{document}

\title{Optical Properties of a $\theta$-Vacuum}
\author{Luis Huerta$^{1,2}$ and Jorge Zanelli$^{3,4}$\\
$^1$  {\small \emph{Departamento de Ciencias Aplicadas, Facultad de Ingenier\'{\i}a, Universidad de Talca}}\\
$^2$  {\small \emph{P4-Center for Research and Applications in Plasma Physics and Pulsed Power Technology, www.pppp.cl, Chile.}}\\
$^3$ {\small \emph{Centro de Estudios Cient\'\i ficos (CECs), Arturo Prat 514, Valdivia, Chile.}} \\
$^4$  {\small \emph{Universidad Andr\'es Bello, Av. Rep\'ublica 440, Santiago, Chile.}}\\
{\small \texttt{lhuerta@utalca.cl, z@cecs.cl}}}
\maketitle

\begin{abstract}
Chern-Simons (CS) forms generalize the minimal coupling between gauge potentials and point charges, to sources represented by charged extended objects (branes). The simplest example of such a CS-brane coupling is a domain wall coupled to the electromagnetic CS three-form. This describes a topologically charged interface where the CS form $AdA$ is supported, separating two three-dimensional spatial regions in 3+1 spacetime. Electrodynamics at either side of the brane is described by the same Maxwell's equations, but those two regions have different vacua, characterized by a different value of the $\theta$ parameter multiplying the Pontryagin form $F \wedge F$. The $\theta$-term is the abelian version of the concept introduced by 't Hooft for the resolution of the $U(1)$ problem in QCD. We point out that CS-generalized classical electrodynamics shows new phenomena when two neighboring regions with different $\theta$-vacua are present. These topological effects result from surface effects induced by the boundary and we explore the consequences of such boundary effects for the propagation of the electromagnetic field in Maxwell theory. Several features, including optical and electrostatic/magnetostatic responses, which may be observable in condensed matter systems, like topological insulators, are discussed.
\end{abstract}

\section{Introduction}

Chern-Simons (\textbf{CS}) forms were accidentally found in mathematics in an attempt to obtain a combinatorial formula for the Pontryagin invariant in four dimensions. The attempt failed, as the authors confessed, ``by the emergence of a boundary term which did not yield to a simple combinatorial analysis." \cite{Chern-Simons}. It turns out that that annoying boundary term has found wide applications in physics, providing Lagrangians for gauge field theories \cite{DJT}, including  three-dimensional gravity \cite{Achucarro-Townsend,Witten88}, and applications to condensed matter physics such as the quantum Hall effect \cite{Zhang,Balachandran}. 

CS theories have a number of remarkable  features besides the fact that they are gauge systems, which makes them interesting as dynamical models for quantum field theories and even potential candidates for gravity and supergravity in higher dimensions (see, e.g., \cite{Z-2005}). The topological nature of the CS forms is reflected in the fact that the CS Lagrangians have no free parameters, no dimensionful coupling constants and therefore no adjustable parameters that run under renormalization. The lagrangian does not require a metric, and the multiplicative constant in front of the action, $k$, can take only quantized values \cite{Witten:1983tw}.

It has been noticed that these functionals have another use: they provide consistent, gauge-invariant couplings between (non-)abelian gauge potentials and extended sources (branes) \cite{Z-2007,Edelstein:2008ry}. The fundamental feature that makes this possible is that under a gauge transformation, the CS forms are not invariant, but quasi-invariant: $\mathcal{C} \rightarrow \mathcal{C}+d($something$)$. Although this might seem as a triviality, it is the key feature that allows coupling the electromagnetic potential $A=A_{\mu}dx^{\mu}$ to a conserved electric current. In fact, retrospect that this can be viewed as a consequence of the fact that the one-form $A$ is itself a CS form. So, it is ironic that physicists had been coupling CS forms to sources for more than a century before mathematicians stumbled upon them, although the recognition of this fact only took place recently.

The CS coupling works in complete analogy with the minimal coupling between the electromagnetic field and a point particle (\textit{0-brane}) of charge $e$,
\begin{equation}
I_{\Gamma^1}=e\int_{\Gamma^1} A_{\mu}(z) dz^{\mu}=e\int_{\Gamma^1} A \,,
\label{EM}
\end{equation} 
where $\Gamma^1$ is the worldline describing the history of the point particle. Similarly, the coupling of a $2p$-brane and a $(2p+1)$-CS form is defined as
\begin{equation}
I_{\Gamma^{2p+1}}=e\int_{\Gamma^{2p+1}} \mathcal{C}_{2p+1}(\mathbf{A}) \,,
\label{CScoupling}
\end{equation} 
where $\mathcal{C}_{2p+1} = Tr[\mathbf{A}\wedge (d\mathbf{A})^p + \alpha_1 \mathbf{A}^3\wedge (d\mathbf{A})^{p-1} + \cdots \alpha_{p}\mathbf{A}^{2p+1}]$ is a  $(2p+1)$ CS form; the coefficients $\alpha_k$ are fixed rational numbers (see, e.g., \cite{Nakahara}), and $\Gamma^{2p+1}$ is the worldvolume embedded in spacetime, swept by the $2p$-brane in its time evolution . That $\mathcal{C}_{2p+1}$ transforms by a closed form under a gauge transformation, $\mathbf{A}\rightarrow \mathbf{A}'=g^{-1}\mathbf{A}g+g^{-1}dg$, is far from obvious. This is a consequence of the relation between CS forms and characteristic classes, such as the Pontryagin and the Euler topological invariants, $P_{2n}=Tr[\mathbf{F}^n]$, where $\mathbf{F}=d\mathbf{A}+\mathbf{A}^2$ is the field strength. The fact that the characteristic classes are closed ($dP_{2n}=0$), implies that they can be locally written as the derivative of some ($2n-1$)-forms (precisely the annoying boundary terms discovered by Chern and Simons),
\begin{equation}
P_{2n}=d\mathcal{C}_{2n-1}.
\label{P=dC}
\end{equation}
Since the CS forms are odd, the couplings of the form (\ref{CScoupling}) are appropriate for even-dimensional branes that sweep odd-dimensional histories in spacetime ($\Gamma^{2p+1}$).

From (\ref{P=dC}), it can be easily seen that under an infinitesimal gauge transformation, $\mathcal{C}$ must change by a boundary term (exact form). Physically, this is the crucial feature that guarantees the consistency and gauge invariance of the coupling to a conserved source. In electromagnetism, the relation between the quasi-invariance of the vector potential and the conservation of charge can be seen writing (\ref{EM}) as
\begin{equation}
I_{\Gamma^1}=\int_{M^D} A\wedge *j, 
\end{equation}
where $*j=e\delta(\Gamma^1)d\xi^1 \wedge \cdots d\xi^{D-1}$ is the dual of the current density, with support on the worldline of the point charge. Under a gauge transformation $\delta A = d\Omega$, $\delta A = d\Omega$,
\begin{equation}
\delta[A \wedge *j]=d\Omega \wedge *j= d[\Omega \wedge *j]- \Omega d*j \, ,
\end{equation}
and therefore, gauge invariance follows from current conservation, $d*j \sim \partial_{\mu} j^{\mu}=0$. In the same manner, the CS coupling between an abelian connection and a conserved current,
\begin{equation}
\delta[\mathcal{C} \wedge *j]=d\Omega \wedge *j= d[\Omega \wedge *j]- \Omega d*j.
\end{equation}
Hence, coupling a CS form to a conserved source guarantees that this variation is a boundary term. 

Couplings of this sort have been considered in the past in various settings, including supergravity \cite{Mora}, in 2+1 AdS gravity \cite{Miskovic:2009uz}, and higher dimensions \cite{Miskovic:2009dd,Edelstein:2010sx,Edelstein:2010sh,EGMZ}. In our four-dimensional spacetime there are two types of branes that can be coupled in this manner: 0-branes (point particles) and 2-branes (ordinary two-dimensional surfaces evolving in spacetime). We are quite familiar with the first type of objects in dynamical theories with point sources, such as electrodynamics and gravitation. 

It is the goal of this paper to explore the second type of coupling in the simplest possible setting in four dimensions, that is, the interaction between a charged 2-brane and an abelian connection. This brane-CS coupling may give rise to interesting observable effects in the propagation of electromagnetic waves at the interface between regions of space characterized by different values of a parameter $\theta\in [0,2\pi]$, the angle that multiplies the Pontryagin density, as originally proposed by 't Hooft in his famous resolution of the $U(1)$ problem in QCD, and is related to the instanton number of the Euclidean theory. This parameter, which may be related to the microscopic nature of special materials, characterizes the topological sector that defines the vacuum state in electromagnetism.

It must be emphasized that the  phenomena described below can be analyzed macroscopically as classical effects, even if their microscopic origin would certainly be quantum. In this spirit, the CS electromagnetic coupling is the consequence of introducing, in a region of space, the Pontryagin invariant, which is the only bilinear for the electromagnetic field that could be added to the classical Maxwell action without spoiling Lorentz and gauge invariance.

A similar construction can be carried out for General Relativity. In that case, one could add to the Einstein-Hilbert action a topological invariant term in a bounded region $\bar{M}$ of spacetime. The resulting theory would have, the usual Einstein equations, describing a metric, torsion-free pseudo-Riemannian manifold almost everywhere. However, the boundary $\partial\bar{M}$ would act as a localized torsion source \cite{Toloza-Zanelli}. Thus, the system analysed here could be viewed as a toy model that describes the analogue of a cosmological model composed of domains with different $\theta$-vacua, separated by domain walls containing torsion.

\section{Electrodynamics of a $\theta$-vacuum in a bounded region\footnote{This section is based on \cite{HZ}.}}
Adding the Pontryagin topological invariant $(\theta/2) F\wedge F$ to the Maxwell action does not affect field equations since its variation is a boundary term that can be dropped under the usual boundary conditions in electromagnetic theory (vanishing fields on the boundary or at infinity). The situation, however, is not the same if the Pontryagin term is present in a bounded domain, surrounded by a larger region where $\theta=0$. In this case, matching conditions, relating fields on both sides of the interface, are not those derived from the purely Maxwellian theory. This results, for example, in a modification of the Casimir energy inside a spherical region characterized by a nonvanishing $\theta$ surrounded by empty space \cite{Canfora:2011fd}. In that case, the boundary spoils the topological invariance of the $\theta$-term, producing a correction to the zero-point energy that is neither negligible nor periodic in $\theta$, as would be expected if the addition were a true topological invariant.

Consider an electromagnetic field in four-dimensional spacetime $M$, where a region $\tilde{M}\subseteq M$ is filled by some material characterized by a parameter $\theta$, such that the action reads
\begin{equation}
I[A]= \frac{1}{2} \int_M F\wedge *F - \frac{\theta}{2} \int_{\tilde{M}\subset M} F\wedge F. \label{theta-action}
\end{equation}
The last term has the form of a topological invariant, but it fails to be topological precisely because it is defined over a bounded region. With the help of the characteristic function 
\begin{equation}
\Theta (x)= \left\{
\begin{array}{l}
\theta , \,  x\in \tilde{M} \\
0 , \,  x \notin \tilde{M}
\end{array}
\right. ,
\end{equation}
the $\theta$-term can also be written as a coupling between the Chern-Simons and a surface current,
\begin{equation}
\int_M \frac{\Theta }{2}F\wedge F=\int_{\partial\tilde{M}} j\wedge A\wedge dA  .  \label{2}
\end{equation}
The surface current is the one-form $\ j=d \Theta=\theta \delta (\Sigma)dz$, where $z$ is the coordinate along outward normal to the surface of $\tilde{M}$, $\Sigma =\partial \tilde{M}$. Since the $\theta$-term is locally exact, the field equations, both inside and outside $\tilde{M}$ are the same as in vacuum. However, this term modifies the behavior of the field \textit{at} the surface $\Sigma $. In fact, varying the action (\ref{theta-action}) yields
\[
d\ast F=j\wedge F , 
\]
or, in more familiar notation,
\begin{equation}
\partial _{\mu }F^{\mu \alpha }=\frac{\theta }{2}\delta (\Sigma )\epsilon^{n\alpha \mu
\nu}F_{\mu \nu }  ,  
\label{3}
\end{equation}
where the index $n$ refers to the normal direction to $\Sigma$. The peculiar feature of the source in (\ref{3}) is that it is proportional to the electromagnetic field itself. Writing (\ref{3}) in coordinates adapted to the surface, one finds\footnote{Here $(\mathbf{E})_{i}=F^{oi}=-F_{0i}$ and $(\mathbf{B})_i=\frac{1}{2}\epsilon_{ijk}F^{jk}$.}
\begin{eqnarray}
\mathbf{\nabla }\cdot \mathbf{E} &\mathbf{=}&\theta \delta (\Sigma )\mathbf{B}\cdot \mathbf{n}
 \label{Gauss} \\
\nabla \times \mathbf{B} -\partial_{t}\mathbf{E}&\mathbf{=}&\theta \delta (\Sigma) \mathbf{E\times n}
 \label{Ampere}
\end{eqnarray}
where $\mathbf{n}$ is the unit normal to $\Sigma $. In the steady state or static case ($\partial_{t}\sim 0$), in the vicinity of the surface $\Sigma$ these equations imply that the normal component of $\mathbf{E}$ and the tangential components of $\mathbf{B}$ are discontinuous, 
\begin{eqnarray}
\lbrack \mathbf{E}_{n}] &=&\theta \mathbf{B}_{n} \label{4} \\
\lbrack \mathbf{B}_{\Vert }] &=&-\theta \mathbf{E}_{\Vert }  \label{5}
\end{eqnarray}
On the other hand, from the identity $dF\equiv 0$ ($\mathbf{\nabla }\cdot \mathbf{B}=0$, $\partial _{t}\mathbf{B+\nabla \times E}=0$), it follows that the normal component of $\mathbf{B}$ and the tangential component of $\mathbf{E}$ must be continuous (in the static limit),
\begin{eqnarray}
\left[ \mathbf{B}_{n}\right]  &=&0  \label{6} \\
\left[ \mathbf{E}_{\Vert }\right]  &=&0  \label{7}
\end{eqnarray}
These continuity conditions imply that the right hand sides of (\ref{4}) and (\ref{5}) are well defined and they represent surface charge and current densities, respectively. 

The phenomenological novelty here is that these sources are given by components of the electromagnetic field itself. The ``surface charge" that appears on the right hand side of (\ref{Gauss}) is proportional to the normal component of the magnetic field (which is well defined on $\Sigma$). This is similar to the behavior of vortices with magnetic flux as carriers of electric charge in superconductors. An immediate consequence of this is that the presence of a magnetic field crossing the surface $\Sigma$ is sufficient to generate an electric field, even in the absence of free electric charges. For instance, a monopole of magnetic charge $g$ surrounded by a spherical region in a $\theta$-vacuum would seem electrically charged for an exterior observer with charge $q=g\theta$, as in the so-called Witten effect \cite{Witten:1979ey}.

On the other hand, the components of the electric field tangent to $\Sigma$ act as surface currents that induce a magnetic field. For example, a static sphere of this $\theta$-material, immersed in a uniform electric field, would respond by generating a magnetic field identical to that of a spinning sphere covered by a uniform charge density.

In these effects, the interface partially transforms electric and magnetic fields into each other, a particular form of  duality transformation \cite{BGH}.  This form of  duality transformation has also potentially observable consequences in the transmission of electromagnetic waves as discussed in the next section.

\section{Electromagnetic waves propagation across $\theta$ boundary}

Wave propagation in both media $M$ and $\tilde{M}$ is governed by Maxwell's equations. However, since the boundary conditions depend on $\theta$, the propagation of electromagnetic waves  across the interface $\partial \tilde{M}$ is necessarily affected, and as a consequence both the reflected and refracted waves experience changes in polarization. On the other hand, the standard geometric laws for reflection and refraction of the wave vectors at the interface hold, since those relations are independent of the polarization plane.

In order to isolate the problem of the $\theta$-interface from other optical effects, let us consider the spacetime regions $M$ and $\tilde{M}$ with unit relative permitivity and permeability, $\epsilon/\epsilon_0=1$, $\mu/\mu_0=1$ (but, $\theta\neq 0$). Assuming an incoming electromagnetic wave of the form
\begin{equation}
\left\lbrace
\begin{array}{c}
\mathbf{E}\\
\mathbf{B}
\end{array}
\right\rbrace
= \left\lbrace\begin{array}{c}
\mathbf{E_0}\\
\mathbf{B_0}
\end{array}
\right\rbrace
e^{i(\omega t - \mathbf{k}\cdot \mathbf{r})}.
\end{equation}
impinging on the surface of $\tilde{M}$. In general, there will be a reflected wave and since in this case the refraction index is one, the transmitted wave emerges with a refraction angle equal to the incidence angle.

The amplitudes and polarization vectors of the reflected and transmitted waves are obtained following the  procedure leading to standard Fresnel equations.  As usual, the electric and magnetic fields can be decomposed into their parallel (\textit{p}-wave, with subscript $\parallel$) and perpendicular (\textit{s}-wave, with subscript $\perp$) components to the plane of incidence, which is defined as perpendicular to the interface $\Sigma$, and containing the direction of propagation. From Maxwell's equations, $\mathbf{B} = \mathbf{k} \times \mathbf{E}$ (with $c=1$), so it is sufficient to write the equations for the electric field. Using subscripts $i$, $r$ and $t$ for the incident, reflected and transmitted (refracted) waves, respectively, we define the relative amplitudes $e_{i\parallel}\equiv E_{i\parallel}/E_i , \qquad e_{r\parallel}\equiv E_{r\parallel}/E_i ,\qquad e_{t\parallel}\equiv E_{t\parallel}/E_i$ (and similar expressions for the $\perp$ components), where $E_i\equiv |\mathbf {E}|=(E_{i\parallel}^2 + E_{i\perp}^2)^{1/2}$. Applying the boundary conditions at the interface, one obtains
\begin{eqnarray}
(e_{r\parallel},e_{r\perp}) = \frac{-\theta}{4 + \theta^2} (\theta e_{i\parallel} + 2 e_{i\perp}, 2 e_{i\parallel} - \theta e_{i\perp}), \\
(e_{t\parallel}, e_{t\perp})  =  \frac{2}{4 + \theta^2} (2e_{i\parallel} - \theta e_{i\perp}, \theta e_{i\parallel} + 2e_{i\perp}),
\end{eqnarray}
for the relative amplitudes of the reflected and transmitted waves. These results imply that the polarization plane of these waves are rotated relative to the polarization plane of the incident wave. If the polarization angle for the incident wave with respect to the plane of incidence is $\alpha_i$, the polarization planes of the reflected and transmitted waves are rotated, respectively, by
\begin{eqnarray}
\triangle \alpha_r  \equiv  \alpha_r - \alpha_i &=& \arctan \left(-\frac{2}{\theta}\right) \\
\triangle \alpha_t  \equiv  \alpha_t - \alpha_i &=& \arctan \left(\frac{\theta}{2}\right)
\end{eqnarray}
(we have assumed $\theta\neq 0$). Thus, independently of the incident polarization plane, $|\triangle \alpha_r - \triangle \alpha_t|= \pi/2$, that is, reflected and refracted waves polarization are perpendicular to each other. This follows from the boundary conditions that mix the electric and magnetic fields. We see that a fully $p$-polarized reflected wave and a fully $s$-polarized refracted wave simultaneously appear for an incident wave polarization angle given by $\tan \alpha_i = 2/\theta$, measured respect to the incident wave. It might seem odd that for $\theta\rightarrow 0$, the reflected wave would show a very strong rotation (close to $\pi/2$), while we expect no effect for $\theta=0$. The paradox is resolved by observing that although the rotation angle can be very large, the amplitude of the reflected wave approaches zero as well.

For $\theta \neq 0$ there is always a reflected wave, and therefore transmittance of the interface is less than 1. In fact, the reflectance $R$ and transmittance $T$ of the boundary surface are given by
\begin{eqnarray}
R  =  e_{r\parallel}^2 + e_{r\perp}^2 = \frac{\theta^2}{4 + \theta^2} \\
T  =  e_{t\parallel}^2 + e_{t\perp}^2 = \frac{4}{4 + \theta^2}
\end{eqnarray}

Therefore, although the $\theta$-term does not influence the local dynamics, the interface does. In particular, for large $\theta$, reflectance approaches unity and the boundary becomes a perfect reflector for electromagnetic waves.  

\section{Discussion}

Our results show that a compact region in spacetime with a $\theta\neq 0$ could be detectable by the interaction of electromagnetic waves with the boundary. Besides reflection, there is a rotation of the polarization plane that is independent of the polarization of the incident wave and of the angle between the wave vector and the surface. This rotation is analogous to the magneto-optical  Kerr effect \cite{Kerr}, in which the wave reflected off a magnetic material has the polarization plane rotated with respect to the incident wave. In the Kerr effect, however, the rotation angle depends on the relative orientation between the magnetization of the material and the wave vector. Another situation where the polarization plane is rotated is the Faraday effect, where the polarization plane of a transmitted wave is rotated in an optical medium in the presence of a magnetic field. In both cases the rotation is produced by off-diagonal components of the dielectric tensor, and can therefore be attributed to the detailed microscopic interaction between the medium and the electromagnetic wave.

The effect we discussed here is related to the one that occurs in topological insulators \cite{gruco11}, in which an effective $\theta$ term is introduced as a effective macroscopic parameter to account for the quantum properties of those systems, whereas here the $\theta$ parameter is understood as a property of the vacuum. The presence of a $\theta$-term is a natural extension of classical electrodynamics, that respects the fundamental properties of the theory, namely gauge invariance and Lorentz symmetry of the vacuum. If the boundary effects are not taken into account, regions with $\theta\neq 0$ would be indistinguishable from the ordinary vacuum space.

The possibility of introducing a $\theta$-term in a gauge theory follows from the existence the characteristic classes in fiber bundles. These topological invariants do not change the local behavior of the dynamical fields. For example, the speed of light is the same in regions with different values of $\theta$, which is why there is not refraction at the interfaces. This gives rise to an interesting cosmological possibility if the universe consisted of a number of regions with different values of $\theta$, which were initially causally disconnected. As the universe expands these regions might have grown to touch each other. A remnant of that scenario today would be a universe divided into distinct domains, like a ferromagnet. An interesting feature of this scenario is that interfaces between regions of different $\theta$ could be detectable by the interaction between electromagnetic waves and the boundaries.

Topological densities obeying relations like (\ref{P=dC}) exist for any gauge theory, be it QCD or gravity. In QCD, the existence of different $\theta$ vacua can lead to observable effects, like a shift in the zero-point energy if the gluons are confined to a region $B$ (bag), where $\theta\neq 0$ surrounded by a $\theta=0$ vacuum \cite{Canfora:2011fd}. In gravitation there are two characteristic classes in four dimensions that can have a similar effect, the Pontryagin invariant, $P_4=R^a_{\;b}\wedge R^b_{\;a}$, and the Euler density, $\mathcal{E}_4=\epsilon_{abcd}R^{ab}\wedge R^{cd}$. To each of these invariants a similar phenomenon could be associated. Although the Einstein equations would not be affected, new phenomena can be expected to arise at the interfaces between regions with different $\theta$s. In particular, the boundaries would act like sources of curvature and torsion. Again, it must be stressed that such ``sources" are not new forms of matter, but they are produced by the geometry itself.

The effects produced by the $\theta$ terms in different regions highlights the fact that our naive understanding of sources for the electromagnetic or gravitational fields, as produced by some forms of matter, may be too narrow. The effect of a distribution of charges can be mimicked by a discontinuity in the definition of vacuum state. We expect these and other classical consequences of the introduction of a $\theta$ vacuum will be repeated in other settings.

============================
\section*{\Large Acknowledgments}

This work was supported by Fondecyt grants \# 1110102, 1100328, 1100755, and by Conicyt grant \textit{Southern Theoretical Physics Laboratory, ACT-91}.  The P4-Center for Research and Applications in Plasma Physics and Pulsed Power Technology is partially supported by Comisi\'{o}n Chilena de Energ\'{\i}a Nuclear. The Centro de Estudios Cient\'{\i}ficos (CECS) is funded by the Chilean Government through the Centers of Excellence Base Financing Program of Conicyt.


\begin{thebibliography}{99}
\bibitem{Chern-Simons} S.~S.~Chern and J.~Simons, \emph{Characteristic forms and geometric invariants}, Annals Math. \textbf{99}, 48-69 (1974).
\bibitem{DJT} S.~Deser, R.~Jackiw and S.~Templeton, \emph{Three dimensional massive gauge theories}, Phys. Rev.Lett. \textbf{48}, 975 (1983); \emph{Topologically massive gauge theory}. Ann. Phys. NY \textbf{140}, 372 (1984).
\bibitem{Achucarro-Townsend}  A.~Ach\'ucarro and P.~K.~Townsend,  \emph{A Chern-Simons action for three-dimensional anti-De Sitter supergravity Theories}, Phys. Lett.  {\bf B180}, 89 (1986).
\bibitem{Witten88} E.~Witten, \emph{(2+1)-Dimensional gravity as an exactly soluble system}, Nucl. Phys. {\bf B311} (1988) 46.
\bibitem{Zhang} S.~C. ~Zhang, T.~H.~Hansson and S.~Kivelson, Phys. \ Rev.\ Lett.\ {\bf 62}, 82 (1989);
S.~C.~Zhang, Int.\ J.\ Mod.\ Phys.\ {\bf B6}, 25 (1992); X.-L.~Qi, T.~Hughes, S.~C.~Zhang, Phys. Rev. {\bf B78}, 195424 (2008).
\bibitem{Balachandran} A.~P.~Balachandran, L.~Chandar and B.~Sathiapalan, \emph{Chern-Simons duality and the quantum Hall effect}, Int.\ J.\ Mod.\ Phys.\  A {\bf 11} (1996) 3587 [arXiv:hep-th/9509019].
\bibitem{Z-2005} J.~Zanelli, \emph{Lecture notes on Chern-Simons (super-)gravities}. [arXiv:hep-th/0502193].
\bibitem{Witten:1983tw} E.~Witten, \emph{Global Aspects of Current Algebra}, Nucl.\ Phys.\  {\bf B223 } (1983) 422-432.
\bibitem{Z-2007} J.~Zanelli, \emph{Uses of Chern-Simons actions}, Talk given at Ten Years of AdS/CFT: A Workshop Celebrating the Tenth Anniversary of the Maldacena Conjecture, Buenos Aires, Argentina, Dec. 2007, AIP Conf.\ Proc.\  {\bf 1031}, 115 (2008)[arXiv:0805.1778 [hep-th]].
\bibitem{Edelstein:2008ry} J.~D.~Edelstein and J.~Zanelli, \emph{Sources for Chern-Simons theories},in  Quantum Mechanics of Fundamental Systems: The Quest for Beauty and Simplicity, (Springer, New York, 2009), p. 117. [arXiv:0807.4217 (hep-th)].
\bibitem{Nakahara} M.~Nakahara, \emph{Geometry, Topology and Physics},  (Institute of Physics, London, 2003).
\bibitem{Mora} P.~Mora and H.~Nishino, \emph{Fundamental extended objects for Chern-Simons supergravity}, Phys. Lett. B \textbf{482}, 222 (2000) [arXiv:hep-th/0002077].
P.~Mora, \emph{Chern-Simons supersymmetric branes}, Nucl. Phys. B \textbf{594}, 229 (2001) [arXiv:hep-th/0008180].
\bibitem{Miskovic:2009uz} O.~Mi\v{s}kovi\'c and J.~Zanelli, \emph{On the negative spectrum of the 2+1 black hole}, Phys.\ Rev.\  D {\bf 79}, 105011 (2009). [arXiv:0904.0475 [hep-th]].
\bibitem{Miskovic:2009dd} O.~Mi\v{s}kovi\'c and J.~Zanelli, \emph{Couplings between Chern-Simons gravities and 2p-branes}, Phys.\ Rev.\  D {\bf 80}, 044003 (2009) [arXiv:0901.0737 [hep-th]].
\bibitem{Edelstein:2010sx} J.~D.~Edelstein, A.~Garbarz, O.~Mi\v{s}kovi\'c and J.~Zanelli, \emph{Stable p-branes in Chern-Simons AdS supergravities}, Phys.\ Rev.\  D {\bf 82}, 044053 (2010) [arXiv:1006.3753 [hep-th]].
\bibitem{Edelstein:2010sh} J.~D.~Edelstein, A.~Garbarz, O.~Mi\v{s}kovi\'c and J.~Zanelli, \emph{Naked Singularities, Topological Defects and Brane Couplings} Int.\ J.\ Mod.\ Phys.\  D {\bf 20}, 839 (2011) [arXiv:1009.4418 [hep-th]].
\bibitem{EGMZ} J.~D.~Edelstein, A.~Garbarz, O.~Mi\v{s}kovi\'c and J.~Zanelli, \emph{Geometry and stability of spinning branes in AdS gravity}, Phys. Rev. D {\bf 84} (in press) [arXiv:1108.3523 (hep-th)].
\bibitem{U1} G.~'t Hooft, \emph{How Instantons Solve the U(1) Problem}, Phys.\ Rept.\  {\bf 142}, 357-387 (1986).
\bibitem{Toloza-Zanelli} A.~Toloza and J.~Zanelli, (in preparation).
\bibitem{HZ} L.~Huerta and J.~Zanelli, \emph{Introduction to Chern-Simons Theories}, Proc. of Sci., ICFI2010 (2010)
004.
\bibitem{Canfora:2011fd} F.~Canfora, L.~Rosa and J.~Zanelli, \emph{Theta term in a bounded region}, Phys. Rev. D {\bf 84} (in press) [arXiv:1105.2490 [hep-th]].
\bibitem{Witten:1979ey} E.~Witten, \emph{Dyons of Charge $e\theta/2 \pi$}, Phys.\ Lett.\  B {\bf 86}, 283 (1979).
\bibitem{BGH} C.~Bunster, A.~Gomberoff and M.~Henneaux, \emph{Extended Charged Events and Chern-Simons Couplings}. [arXiv:1108.1759 (hep-th)].
\bibitem{Kerr} J.~Kerr, Phil.\ Mag.\ {\bf 3}, 321 (1877); J.~Kerr, Phil.\ Mag.\ {\bf 5}, 161 (1878). See also P.~Weinberger, \emph{John Kerr and his effects found in 1877 and 1878}, Phil.\ Mag.\ Lett.\ {\bf 88}, 897 (2008).
\bibitem{gruco11} A. G. Grushin, and A. Cortijo, \emph{Tunable Casimir Repulsion  with three-dimensional Topological Insulators}, Phys.\ Rev. \ Lett. \textbf{106}, 020403 (2011); A.~G.~Grushin, P.~Rodriguez-Lopez and A.~Cortijo, \emph{Effect of finite temperature and uniaxial anisotropy on the Casimir effect with three-dimensional topological insulators} arXiv:1102.0455[cond-mat.mtrl-sci] (unpublished)

\end{thebibliography}
\end{document}